# Optimisation models for the day-ahead energy and reserve scheduling of a hybrid wind-battery virtual power plant

Daniel Fernández-Muñoz[a,*] (Researcher), Juan I. Pérez-Díaz[b] (Coordinator)

[a]Department of Physics Electronics, Electrical Engineering and Applied Physics, ETSIT, Universidad Politécnica de Madrid, Avenida Complutense 30, 28040, Madrid, Spain

[b]Department of Hydraulic, Energy and Environmental Engineering, ETSICCP, Universidad Politécnica de Madrid, C/ Profesor Aranguren 3, 28040, Madrid, Spain

ABSTRACT

This work presents a suite of two optimisation models for the short-term scheduling and redispatch of a virtual power plant (VPP) composed of a wind farm and a Li-ion battery, that participates in the day-ahead energy and secondary regulation reserve markets of the Iberian electricity market. First, a two-stage stochastic mixed-integer linear programming model is used to obtain the VPP's generation and reserve schedule and the opportunity cost of the energy stored in the battery. The model has an hourly resolution and a look-ahead period beyond the markets' scheduling horizon and considers the hourly battery degradation costs as a function of both the depth of discharge and the discharge rate. Different strategies are evaluated to forecast the real-time use of the committed secondary regulation reserves. Second, a deterministic MILP model is used to determine the redispatch of the VPP using as input the generation and reserve schedule and the VPP's storage opportunity cost provided by the former model and is executed on an hourly rolling basis. The results obtained show that the proposed models are effective for the short-term scheduling and redispatch of the VPP used with a low computational time, making them tractable and practical for their daily use.

## 1. Introduction

### 1.1. Motivation

In the last decade, a number of initiatives worldwide have been launched with the aim of reducing the greenhouse gas emissions of the electricity system, the so-called power system decarbonisation [1]. The European Union has set an ambitious target of at least 32% of renewable energy share for 2030. This target might be revised upward during 2023 according to the European Directive 2018/2001 [2]. The energy transition required to meet such a target is opening opportunities for distributed energy resources (DER). A Virtual Power Plant (VPP) is a "representation" of a DER portfolio to participate in the wholesale market, i.e., an aggregation of different DERs, usually from the same geographical location, managed in a coordinated manner [3, 4].

Spain is not indifferent to this energy transition and in the last years, a number of regulatory changes have been implemented, such as the RD 244/2019 [5], which regulates the administrative, technical, and economic conditions for self-consumption, and the RDL 233/2020 [6], which introduces the figure of renewable energy communities as a subject that can participate in the activity of electricity production. Recently, there has been a public consultation regarding some modifications of the operating procedures of the Spanish's transmission system operator, that will allow the aggregation of generation, storage, and/or demand resources, to offer balancing services, in accordance with [7].

VPPs will therefore be soon able to improve their economic feasibility thanks to the participation in the reserve markets.

### 1.2. Literature review

The interest of the scientific community in the study of VPPs, generally comprising variable renewable generation technologies together with storage technologies, has grown remarkably in the last years. Several authors have presented optimisation models for the short-term scheduling of VPPs comprising variable renewable generation technologies and energy storage participating in the day-ahead electricity and reserve markets [8–10] and, in other cases, of VPPs that also consider conventional generation technologies [11–13].

The authors of [12] proposed a deterministic mixed-integer non-linear programming model for the short-term scheduling of a VPP comprising wind, solar photovoltaic (PV) and conventional generation together with a battery energy storage system (BESS). A deterministic mixed-integer linear programming (MILP) model is proposed in [14] for the short-term scheduling of a VPP comprising wind and PV generation and energy storage, which participates only in the day-ahead electricity market. A similar approach can be found in [15]. The authors of [8] present a stochastic MILP model to obtain the generation schedule of a VPP comprising wind and PV generation, a pump station and a BESS, that participates in the day-ahead electricity and reserve markets. The authors of [8] compare the results obtained when the VPP is scheduled in a coordinated manner versus an uncoordinated one, showing better results for the former. A multi-stage stochastic MILP model is presented in [9] for the short-term scheduling of a VPP comprising wind generation and a BESS that participates in the day-ahead electricity and reserve markets of the Iberian electricity

---

*Corresponding author
daniel.fernandezm@upm.es (D. Fernández-Muñoz)
ORCID(s): 0000-0003-1789-3151 (D. Fernández-Muñoz)

# Nomenclature

*Indexes and sets*

$\Omega_{t,s}$  Set of decision variables of the GRM model
$\Psi_{t,s}$  Set of decision variables of the GRM model's first stage
$j, J$  set of battery degradation cost curves [1,10]
$s, S$  stochastic scenarios of the hourly wind energy generation ($s'$ is a mirror index of $s$)
$t, T$  hourly periods of the planning period
$T^1$  hourly periods corresponding to the first stage [1,24]

*GRM model parameters*

$\eta$  total round-trip battery efficiency [0.86]
$\mu_t^{up}/\mu_t^{dw}$  percentage of the secondary regulation upward/downward reserve that will be required in real-time in period $t$
$\overline{c}_j^b$  maximum battery degradation cost [€]
$\overline{p}^b$  battery output/input power capacity [MW]
$\overline{p}^w$  wind farm capacity [33 MW]
$\pi_t^{\Delta^{up}}/\pi_t^{\Delta^{dw}}$  upward/downward energy imbalance market price in period $t$ [€/MWh]
$\pi_t^D$  day-ahead electricity market price in period $t$ [€/MWh]
$\pi_t^{S,up}/\pi_t^{S,dw}$  upward/downward secondary regulation energy market price in period $t$ [€/MWh]
$\pi_t^S$  secondary regulation reserve market price in period $t$ [€/MWh]
$\rho_t$  ratio between the upward and the total secondary regulation reserve in period $t$, set by the TSO in advance
$\underline{c}_j^b$  minimum battery degradation cost corresponding to curve $j$ [€]
$\underline{soc}/\overline{soc}$  minimum/maximum state of charge [MWh]
$L_j^b$  minimum initial battery state of charge corresponding to curve $j$ [MWh]
$r_t^{w,pk}/r_t^{w,ik}$  wind generation forecast in period $t$ with perfect/imperfect knowledge [MWh]
$r_{t,s}^w$  wind generation forecast in period $t$ and scenario $s$ [MWh]
$s_j^b$  battery degradation cost slope corresponding to curve $j$ [€/dcr]

*GRM model variables*

$\Delta_{t,s}^{up}/\Delta_{t,s}^{dw}$  expected upward/downward wind energy imbalance in period $t$ and scenario $s$ [MWh]
$c_{t,s}^b$  battery degradation cost in period $t$ and scenario $s$ [€]
$dcr_{t,s}$  battery discharge rate in period $t$ and scenario $s$ [0,1]
$e_{t,s}^{D,c}/e_{t,s}^{D,d}$  battery energy charge/discharge in the day-ahead energy market in period $t$ and scenario $s$ [MWh]
$e_{t,s}^{S,up}/e_{t,s}^{S,dw}$  secondary regulation upward/downward reserve energy requested in real-time by the TSO in period $t$ and scenario $s$ [MWh]
$g_{t,s}^{S,up}/g_{t,s}^{S,dw}$  secondary regulation upward/downward reserve in period $t$ and scenario $s$ [MW]
$gr_{t,s}^{S,up}/gr_{t,s}^{S,dw}$  secondary regulation upward/downward reserve set aside to cope with wind energy imbalances in period $t$ and scenario $s$ [MWh]
$p_{t,s}^b$  battery discharge/charge output/input power in the day-ahead electricity market in the period $t$ and scenario $s$ [MWh] (positive/negative value indicates discharge/charge mode, respectively)
$p_{t,s}^w$  wind energy generation in period $t$ and scenario $s$ [MWh]
$soc_{t,s}$  battery state of charge in period $t$ and scenario $s$ [MWh]

*GRM model integer variables*

$D_j$  0 if $j=1$; $j-1$ for battery degradation cost curve $j$
$u_{t,s}$  1 if $dcr_{t,s} > 0$ in period $t$ and scenario $s$; 0 otherwise
$y_{t,s}^a$  1 if $e_{t,s}^{D,d} > 0$; 0 otherwise
$y_{t,s}^b$  1 if $\Delta_{t,s}^{up} > 0$; 0 otherwise
$y_{t,s}^c$  1 if $e_{t,s}^{S,up} > 0$; 0 otherwise

*EDM model parameters*

$k^S$  penalty for secondary regulation reserve deficit [1.5]
$soc^v$  VPP's storage opportunity cost [€/MWh]

*EDM model variables*

$\Delta_t^{soc}$  battery's state of charge deviation between the value given by GRM model and the one given by EDM model in period $t$ [MWh]
$\Delta_{D,t}^{vpp^{up}}/\Delta_{D,t}^{vpp^{dw}}$  VPP's upward/downward energy imbalance with respect to the energy scheduled in the day-ahead electricity market by the GRM model in period $t$ [MWh]
$\Delta_t^{vpp^{up}}/\Delta_t^{vpp^{dw}}$  VPP's upward/downward energy imbalance with respect to the energy scheduled in the day-ahead electricity market by the GRM model and the secondary regulation energy required in real-time in period $t$ [MWh]
$\epsilon_t^{g^{S,up}}/\epsilon_t^{g^{S,dw}}$  upward/downward secondary regulation reserve imbalance with respect to the reserve scheduled by the GRM model in period $t$ [MWh].
$\epsilon_t^{S,up}/\epsilon_t^{S,dw}$  upward/downward secondary regulation energy imbalance with respect to the upward/downward secondary regulation energy required in real-time by the TSO in period $t$ [MWh].

*EDM model integer variables*

$y_t^d$  1 if $\Delta_{D,t}^{vpp^{up}} > 0$; 0 otherwise
$y_t^e$  1 if $\Delta_t^{vpp^{up}} > 0$; 0 otherwise
$y_t^f$  1 if $\Delta_{D,t}^{vpp^{dw}} > 0$; 0 otherwise

market. The authors of [9] compare the contribution of the BESS when the VPP participates only in the day-ahead electricity market and later, when also participates in the reserve market, with an overall improvement of the results. A similar VPP configuration is used in [10], where a two-stage convex stochastic programming model is proposed for the short-term scheduling in the day-ahead electricity market and reserve market of the Iberian electricity market, considering as well the intraday electricity markets to cope with the deviation with respect to the commitments acquired in the day-ahead electricity market. A two-stage stochastic MILP model is presented in [11] for the short-term scheduling of a VPP comprising wind, PV and conventional generation together with a Lithium-ion (Li-ion) battery, in the day-ahead electricity and reserve markets. The model considers the risk aversion of low-income daily scenarios.

Other works that consider the participation of the VPP only in the day-ahead electricity market can be found in [16], where a stochastic MILP model is proposed for the short-term scheduling of a VPP comprising PV, wind and conventional generation, photovoltaic-thermal panels, combined heat and power generation, a BESS and a boiler. The authors of [13] propose a risk-averse bi-level programming model for the short-term scheduling of a VPP comprising wind generation, residential load, and conventional generation, that participates in the day-ahead electricity market also considering the upward and downward deviations.

In addition, some works found in the literature propose short-term optimisation models for VPPs exclusively composed by storage technologies, such as in [17–19]. The authors of [17] present a deterministic linear programming model for the short-term scheduling of a fleet of electric vehicles in the day-ahead electricity market, with the aim of minimising the cost of charging the electric vehicles. Similarly, the authors of [18] propose a deterministic MILP model for the operation of a BESS aggregator that participates in day-ahead electricity and reserve markets. In [19], a deterministic MILP model is presented to perform a business model analysis of a BESS that participates in the day-ahead electricity market of the Iberian electricity market considering different combinations of power and storage of the BESS, and performing an optimal arbitrage strategy.

It is worth mentioning that almost all energy storage systems considered in the above-mentioned works are battery-based ones. None of these works consider the battery degradation cost as a function of the depth of discharge and the discharge rate as suggested by [20].

As can be deduced from [21], in order to maximise the profit an energy storage device with limited energy storage capacity can obtain in the day-ahead electricity and reserve markets, it is important to formulate the short-term scheduling problem with a look-ahead period beyond the markets horizon. None of the above-mentioned works considers such look-ahead period. Only the works of [9, 10, 14, 18] explicitly consider an end-of-day storage target which is set equal to the initial state of charge. As can be deduced from [21] the use of such a target may have a negative impact on the plant's profit.

Finally, it is worth mentioning that most of the above-mentioned works do not address the redispatch of the considered VPP taking into account the realisation of the uncertainty of the variable renewable generation and of the amount of committed reserves that are required in real-time. Among the aforementioned works, only the authors of [17] study the redispatch of the VPP, in this case an aggregation of EVs, taking into account the realisation of the uncertainty in the EVs' usage. For this purpose, the authors of [17] use an operational management algorithm, aimed to minimise the differences between the actual charge/discharge profile of the EVs and the commitments acquired in the day-ahead electricity and reserve markets.

### 1.3. Literature review conclusions

From the literature review one can conclude that the optimisation models proposed in the literature for the short-term generation and reserve scheduling of VPPs comprising variable renewable generation and BESS have some important flaws:

- the battery degradation costs are either neglected or considered in a too simplified manner,
- the economic value of the stored energy at the end of the considered markets' scheduling horizon is disregarded,
- the redispatch of the VPP as a measure to cope with the realisation of the uncertainty of the renewable generation and of the amount of committed reserves that are required in real-time.

### 1.4. Objective and contribution

The main objective of this work is to present a suite of optimisation models for the short-term scheduling and redispatch of a VPP comprising wind generation and a Li-ion BESS, that participates in the day-ahead electricity and secondary regulation reserve markets of the Spanish power system. The main contributions of this work can be summarised as follows:

- the short-term scheduling model uses a *look-ahead* period to consider the opportunity cost of the stored energy in the BESS at the end of the markets scheduling horizon,
- the models consider the battery degradation cost as a function of both the depth of discharge and the discharge rate,
- the redispatch of the VPP is determined taking into account the realisation of the uncertainty of the wind generation and of the amount of committed reserves that are required in real-time.

### 1.5. Paper organisation

The paper is organised as follows. Section 2 describes the VPP used as a case study. Section 3 presents the mathematical formulation of the short-term scheduling model and of the redispatch model. In Section 4 the results obtained are discussed and finally, Section 5 summarises the main conclusions.

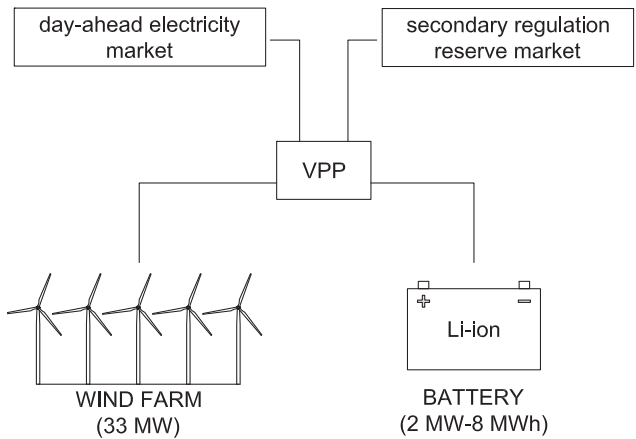

**Figure 1:** Simplified diagram of the VPP and its relation to the Iberian electricity markets considered.

**Table 1**
Parameters of the Li-ion battery considered in the case study.

| Parameter | Adopted value |
|---|---|
| round-trip efficiency ($\eta$) | 86% |
| Maximum state of charge allowed | 100% × $\overline{soc}$ |
| Minimum state of charge allowed | 20% × $\overline{soc}$ |
| Battery cost | 386.4 €/kWh |

## 2. Case study description
### 2.1. Virtual Power Plant description

The VPP used as case study in this work participates as a price-taker in the day-ahead electricity and secondary regulation reserve markets of the Spanish power system. The VPP is composed of a wind farm with a total installed capacity of 33 MW and a Li-ion battery of 2 MW power output/input capacity and a storage capacity of 8 MWh. The VPP offers only the capacity of the battery in the secondary regulation reserve market. A simplified diagram of the VPP considered in this work is shown in Fig. 1. As depicted, the VPP participates in both the day-ahead electricity and secondary regulation reserve markets. The Li-ion battery parameters are taken from [22] and summarised in Table 1.

### 2.2. Data sources

Two different data sources are used as inputs to the optimisation models: a set of forecasts and scenarios is used as input to the short-term scheduling model, and a set of historical data is used as input to the redispatch model. The set used corresponds to 81 representative cases of 2016 selected according to the average value of wind generation (low, medium, high), the variability of wind generation (low, medium, high), the average value of day-ahead electricity market price (low, medium, high) and, the variability of day-ahead electricity market price (low, medium, high), similarly to [23].

Forecasts and scenarios used as input to the short-term scheduling model are obtained as follows: 1) the day-ahead electricity market price is obtained according to the forecast model presented in [24], 2) the secondary regulation reserve market price is obtained according to the forecast model presented in [24], 3) the upward/downward secondary regulation energy market price is calculated in every hour $t$ as the average value in the same hour $t$ across all days in the previous year according to [25], 4) similarly, the upward/downward imbalance price is calculated by adding to the day-ahead electricity market price in every hour $t$ the average difference in the same hour $t$ across all days in the previous year between the imbalance price and the day-ahead electricity market price, 5) different forecasts of the percentage of the committed secondary regulation reserves that are required in real-time (upward and downward) are considered as exposed in Section 4, 6) the wind generation scenarios have been obtained similarly to [23].

The historical market data used as input to the redispatch model have been taken from [26]. Such data are listed as follows: 1) day-ahead electricity market price, 2) upward and downward imbalance price, 3) secondary regulation reserve market price, 4) secondary regulation upward/downward energy market price, 5) percentage of the committed secondary regulation reserves that are required in real-time (upward and downward). In addition, the authors possess a historical time series of the hourly electricity production of the wind farm described in Section 2.1.

## 3. Optimisation models

Two optimisation models are proposed for the short-term scheduling and redispatch of the VPP described in Section 2. First, a two-stage stochastic MILP model with a planning period of $T$ hours and with an hourly resolution (hereinafter referred to as GRM model) is used to obtain the next-day generation and reserve schedule of the VPP as well as the VPP's storage opportunity cost. The GRM model considers the battery degradation cost. Then, both the next-day generation and reserve schedule and the VPP's storage opportunity cost are used as an input to a deterministic hourly redispatch MILP-based model (hereinafter referred to as EDM model), which is executed in an hourly rolling basis during the next day. The EDM model also considers the battery degradation cost.

An outline of the execution sequence of the models is depicted in Fig. 2: first, the GRM model is executed and the next-day generation schedule and reserves are obtained together with the storage opportunity cost of the VPP; then, prior to each hourly period $t$, the EDM model is executed considering the results given by the GRM model for each hourly period $t$ and also the state of charge of the battery resulting from the EDM model applied in the hourly period $t-1$. After twenty-four hours, GRM model is executed again and the rolling basis scheme of the two-step methodology continues. The formulation of both optimisation models is presented below.

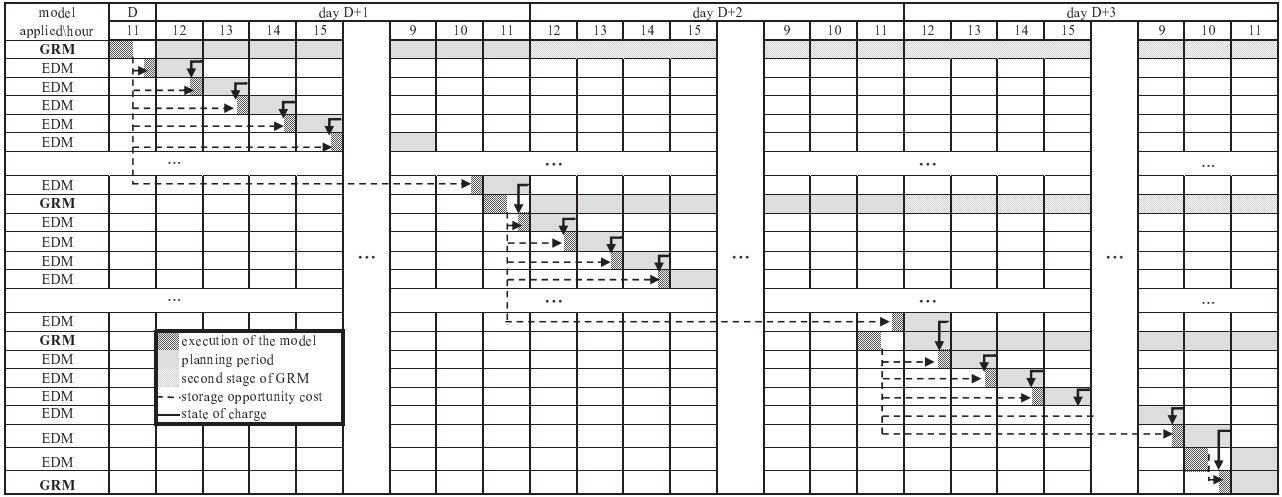

**Figure 2:** Daily and hourly rolling basis scheme of the two-step methodology proposed.

### 3.1. Formulation of GRM model

The GRM model has two decision stages: the first stage ($t \in T^1$) corresponds to the first twenty-four hours of the planning period $T$, the *here and now decisions* period, and the second stage ($t \notin T^1$) extends from hour twenty-five to the end of the planning period $T$. A set $s$ of hourly wind energy scenarios is used to consider the uncertainty of the wind generation for the whole planning period. The scenarios have been obtained similarly to [23]. The GRM model uses forecast values of the following: the hourly prices of the day-ahead electricity market, hourly prices of the secondary regulation reserve and the secondary regulation energy (upward and downward), the hourly imbalance prices, and finally, the hourly percentage of the committed secondary regulation reserves that are required in real-time. The objective function of the GRM model is given by (1).

$$\max_{\Omega_{t,s}} \mathbb{E} \sum_t \left\{ \begin{array}{c} \pi_t^D \cdot (p_{t,s}^w + p_{t,s}^b) + \pi_t^{\Delta^{up}} \cdot \Delta_{t,s}^{up} - \pi_t^{\Delta^{dw}} \cdot \Delta_{t,s}^{dw} \\ + \pi_t^S \cdot (g_{t,s}^{S,up} + g_{t,s}^{S,dw}) \\ + \pi_t^{S,up} \cdot e_{t,s}^{S,up} - \pi_t^{S,dw} \cdot e_{t,s}^{S,dw} - c_{t,s}^b \end{array} \right\} \quad (1)$$

$$\Psi_{t,s} = \Psi_{t,s'} \quad \forall s \neq s', t \in T^1 \quad (2)$$

$$-\overline{p}^b \leq p_{t,s}^b \leq \overline{p}^b \quad \forall s, t \quad (3)$$

$$e_{t,s}^{D,d} - e_{t,s}^{D,c} = p_{t,s}^b \quad \forall s, t \quad (4)$$

$$e_{t,s}^{D,d} \leq \overline{p}^b \cdot y_{t,s}^a \quad \forall s, t \quad (5)$$

$$e_{t,s}^{D,c} \leq \overline{p}^b \cdot (1 - y_{t,s}^a) \quad \forall s, t \quad (6)$$

$$p_{t,s}^w \leq \overline{p}^w \quad \forall s, t \quad (7)$$

$$\Delta_{t,s}^{up} \leq \overline{p}^w \cdot y_{t,s}^b \quad \forall s, t \quad (8)$$

$$\Delta_{t,s}^{dw} \leq \overline{p}^w \cdot (1 - y_{t,s}^b) \quad \forall s, t \quad (9)$$

$$\Delta_{t,s}^{up} - \Delta_{t,s}^{dw} = r_{t,s}^w - p_{t,s}^w + gr_{t,s}^{S,up} - gr_{t,s}^{S,dw} \quad \forall s, t \quad (10)$$

$$gr_{t,s}^{S,up} \leq 2 \cdot \overline{p}^b \cdot (1 - y_{t,s}^b) \quad \forall s, t \quad (11)$$

$$gr_{t,s}^{S,dw} \leq 2 \cdot \overline{p}^b \cdot y_{t,s}^b \quad \forall s, t \quad (12)$$

$$g_{t,s}^{S,up} \leq \begin{cases} \overline{p}^b - p_{t,s}^b - gr_{t,s}^{S,up} \\ 2 \cdot \overline{p}^b \end{cases} \quad \forall s, t \quad (13)$$

$$g_{t,s}^{S,dw} \leq \begin{cases} \overline{p}_{t,s}^b + p^b - gr_{t,s}^{S,dw} \\ 2 \cdot \overline{p}^b \end{cases} \quad \forall s, t \quad (14)$$

$$g_{t,s}^{S,up} = \rho_t \cdot (g_{t,s}^{S,up} + g_{t,s}^{S,dw}) \quad \forall s, t \quad (15)$$

$$e_{t,s}^{S,up} - e_{t,s}^{S,dw} = \mu_t^{up} \cdot g_{t,s}^{S,up} - \mu_t^{dw} \cdot g_{t,s}^{S,dw} \quad \forall\, s,t \quad (16)$$

$$e_{t,s}^{S,up} \leq \overline{p}^b \cdot y_{t,s}^c \quad \forall\, s,t \quad (17)$$

$$e_{t,s}^{S,dw} \leq \overline{p}^b \cdot (1 - y_{t,s}^c) \quad \forall\, s,t \quad (18)$$

$$\begin{aligned} soc_{t,s} = soc_{t-1,s} + \eta \cdot (e_{t,s}^{D,c} + e_{t,s}^{S,dw} + gr_{t,s}^{S,dw}) \\ - (e_{t,s}^{D,d} + e_{t,s}^{S,up} + gr_{t,s}^{S,up}) \quad \forall\, s,t \end{aligned} \quad (19)$$

$$\underline{soc} \leq soc_{t,s} \leq \overline{soc} \quad \forall\, s,t \quad (20)$$

$$dcr_{t,s} \geq \frac{soc_{t-1,s} - soc_{t,s}}{\overline{soc}} \quad \forall\, s,t \quad (21)$$

$$u_{t,s} \geq dcr_{t,s} \quad \forall\, s,t \quad (22)$$

$$\delta_{t,j-1,s} \geq \delta_{t,j,s} \quad \forall\, j \in [2, J),\, s,t \quad (23)$$

$$soc_{t-1,s} \geq L_{j-1}^b \cdot (\delta_{t,j-1,s} - \delta_{t,j,s}) + L_j^b \cdot \delta_{t,j,s} \\ \forall\, j \in [2, J),\, s,t \quad (24)$$

$$\begin{aligned} soc_{t-1,s} \leq \overline{soc} \cdot \delta_{t,j,s} + L_{j-1}^b \cdot (1 - \delta_{t,j-1,s}) \\ + L_j^b \cdot (\delta_{t,j-1,s} - \delta_{t,j,s}) \quad \forall\, j \in [2, J),\, s,t \end{aligned} \quad (25)$$

$$c_{t,s}^b \leq u_{t,s} \cdot \underline{c}_j^b + dcr_{t,s} \cdot s_j^b + \overline{c}_J^b \cdot (\delta_{t,j,s} + \delta_{t,j+1,s}) \\ \forall\, j=1,\, s,t \quad (26)$$

$$c_{t,s}^b \leq u_{t,s} \cdot \underline{c}_j^b + dcr_{t,s} \cdot s_j^b + \overline{c}_J^b \cdot \left[ D_j - \sum_{n=2}^{j-1} \delta_{t,n,s} + \delta_{t,j,s} \right] \\ \forall\, j \in [2, J),\, s,t \quad (27)$$

$$c_{t,s}^b \geq u_{t,s} \cdot \underline{c}_j^b + dcr_{t,s} \cdot s_j^b + \overline{c}_J^b \cdot (\delta_{t,j,s} + \delta_{t,j+1,s}) \\ \forall\, j=1,\, s,t \quad (28)$$

$$c_{t,s}^b \geq u_{t,s} \cdot \underline{c}_j^b + dcr_{t,s} \cdot s_j^b + \overline{c}_J^b \cdot \left[ D_j - \sum_{n=2}^{j-1} \delta_{t,n,s} + \delta_{t,j,s} \right] \\ \forall\, j \in [2, J),\, s,t \quad (29)$$

The first term of (1) represents the expected result in the day-ahead electricity market due to the generation of wind energy and the discharge / charge power output of the battery. The second and third terms refer to the expected positive/negative deviations of the wind energy generation and the corresponding revenue or cost according to the energy imbalance market prices. The fourth term corresponds to the expected revenue in the secondary regulation reserve market, considering that the VPP exclusively offers the battery capacity in this market as previously exposed. The fifth and sixth terms correspond to the expected revenue and cost resulting from the upward and downward secondary regulation energy, respectively. Finally, the last term corresponds to the battery degradation cost.

A set of non-anticipativity constraints (2) is included in the formulation to ensure that the decisions obtained in the first stage ($t \in T^1$) are unique for all wind energy scenarios.

The maximum power output of the battery in discharge or charge mode is given by (3). Equation (4) calculates the energy scheduled for the battery in the day-ahead electricity market, and equations (5)-(6) are used to force the battery energy schedule to be in discharge or charge mode exclusively.

Wind energy is scheduled in the day-ahead electricity market according to (7). The GRM model considers wind energy imbalances ($\Delta_{t,s}^{dw}$ / $\Delta_{t,s}^{up}$) with respect to the wind energy scheduled in the day-ahead electricity market ($p_{t,s}^w$). The binary variable ($y_{t,s}^b$) takes a value of 1 when a positive wind energy imbalance is expected (8) and a value of 0 when a negative wind energy imbalance or no imbalance is expected (9). The GRM model can set aside part of the available secondary regulation reserve ($gr_{t,s}^{S,up}/gr_{t,s}^{S,dw}$) so as to reduce such imbalances (10). The maximum amount of reserve that can be set aside for this purpose is given by equations (11) and (12).

The maximum secondary regulation upward and downward reserve that can be made available in the secondary regulation reserve market is given by equations (13) and (14), respectively. Equation (15) forces the secondary regulation reserve made available in the reserve market to satisfy the ratio $\rho_t$ between the upward and the total secondary regulation reserve set by the transmission system operator (TSO) in advance, similarly to [27]. The expected upward and downward secondary regulation energy is given by (16) taking into account the expected percentage of the committed reserves that will be required in real-time ($\mu_t^{up}/\mu_t^{dw}$). Equations (17) and (18) force the secondary regulation energy to be either upward or downward for each period $t$ and scenario $s$.

The battery state of charge for a period $t$ and scenario $s$ ($soc_{t,s}$) is given by equations (19) and (20). The battery degradation cost considered in the objective function (1) is calculated according to the methodology proposed in [20], as a function of the battery discharge rate ($dcr_{t,s}$) and the initial state of charge of the battery ($soc_{t-1}$). A set $J$ of battery degradation cost curves as a function of the discharge rate has been considered and depicted in Fig. 3. Each curve ($L_j^b$) corresponds to a different value of the initial battery state of charge.

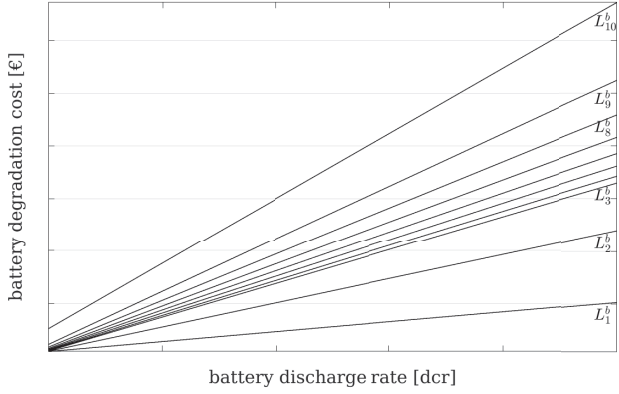

**Figure 3:** Battery degradation cost curves $J$ considered as a function of the initial state of charge and the discharge rate.

The battery discharge rate in period $t$ and scenario $s$ is given by (21). Equation (22) determines if the battery is discharging in period $t$ and scenario $s$. If so, the battery degradation cost is applied. In order to select the battery degradation cost curve that corresponds to the initial state of charge of the battery in period $t$ and scenario $s$, the methodology proposed by the authors of [28] has been used (see equations (23)-(29)).

In summary, the GRM model is defined by the objective function (1) subject to the constraints (2)-(29).

### 3.2. Formulation of EDM model

Before presenting the formulation of EDM model, it is worth mentioning the following:

- The VPP's generation and reserve schedules ($p_{t,1}^w$, $p_{t,1}^b$ and $g_{t,1}^{S,up}$, $g_{t,1}^{S,dw}$) given by the GRM model are considered as obligations for the VPP. Please note that the selection of the stochastic scenario $s=1$ is arbitrary and made for ease of nomenclature, since due to the equation (2), the variables have the same value for a period $t \in T^1$ across all scenarios.

- The EDM model considers the real value of wind generation and the percentage of the committed reserves that are required in real-time. These values are not actually known until the end of each period $t$. However, we find reasonable to assume that they are known at the beginning of each period $t$ with sufficient accuracy.

- The EDM model considers the real value of the cleared prices of the day-ahead electricity and secondary regulation reserve markets. They are disclosed to market participants several hours before the start of the period $t$.

- The upward/downward imbalance and secondary regulation energy market prices are equal to the ones considered in the GRM model. Better forecasts might be available at the beginning of each period $t$. However, we have chosen not to update these price forecasts.

- The symbols used to refer to input data are the same as those used in the description of the GRM model with the superscript $*$.

- The variables of the GRM model that are recalculated by the EDM model are referred to using the same symbol as in the description of the GRM model with the superscript $*$, without considering the wind energy scenario $s$.

- The EDM model considers the penalty dictated in the TSO rules in case the VPP is not able to comply with the above-mentioned obligations.

- The EDM model considers the daily VPP's storage opportunity cost ($soc^v$) that has been calculated by the GRM model.

The EDM model is a deterministic MILP-based redispatch model that aims to maximise the objective function (30) in each hourly period $t \in T^1$.

$$\max_{\Omega_t^*} \left\{ \begin{array}{l} \pi_t^{D*} \cdot (p_{t,1}^w + p_{t,1}^b) + \pi_t^{\Delta up*} \cdot \Delta_t^{vpp^{up}} - \pi_t^{\Delta dw*} \cdot \Delta_t^{vpp^{dw}} \\ + \pi_t^{S*} \cdot (g_{t,1}^{S,up} + g_{t,1}^{S,dw} - k^S \cdot (\epsilon_t^{g^{S,up}} + \epsilon_t^{g^{S,dw}})) \\ + \pi_t^{S,up*} \cdot e_t^{S,up*} - \pi_t^{S,dw*} \cdot e_t^{S,dw*} \\ - c_t^{b*} + \Delta_t^{soc} \cdot soc^v \end{array} \right\} \quad (30)$$

$$p_t^{w*} \leq r_t^{w*} \quad (31)$$

$$\Delta_{D,t}^{vpp^{up}} \leq (\overline{p}^w + \overline{p}^b - p_{t,1}^w - p_{t,1}^b) \cdot y_t^d \quad (32)$$

$$\Delta_{D,t}^{vpp^{dw}} \leq (p_{t,1}^w + p_{t,1}^b + \overline{p}^b) \cdot (1 - y_t^d) \quad (33)$$

$$\Delta_{D,t}^{vpp^{up}} - \Delta_{D,t}^{vpp^{dw}} = p_t^{w*} + p_t^{b*} - p_{t,1}^w - p_{t,1}^b \quad (34)$$

$$g_t^{S,up*} \leq \overline{p}^b - p_t^{b*} \quad (35)$$

$$g_t^{S,dw*} \leq p_t^{b*} + \overline{p}^b \quad (36)$$

$$\epsilon_t^{g^{S,up}} \geq g_{t,1}^{S,up} - g_t^{S,up*} \quad (37)$$

$$\epsilon_t^{g^{S,dw}} \geq g_{t,1}^{S,dw} - g_t^{S,dw*} \quad (38)$$

$$e_t^{S,up*} + \epsilon_t^{S,up} - e_t^{S,dw*} - \epsilon_t^{S,dw} = \mu_t^{up*} \cdot g_{t,1}^{S,up} - \mu_t^{dw*} \cdot g_{t,1}^{S,dw} \quad (39)$$

$$\Delta_{D,t}^{vpp^{dw}} \leq y^f \cdot \left(p_{t,1}^w + p_{t,1}^b + \overline{p}^b\right) \quad (40)$$

$$e_t^{S,up*} \leq \left(1 - y^f\right) \cdot 2 \cdot \overline{p}^b \quad (41)$$

$$\Delta_t^{vpp^{up}} - \Delta_t^{vpp^{dw}} = \Delta_{D,t}^{vpp^{up}} - \Delta_{D,t}^{vpp^{dw}} - \left(\epsilon_t^{S,up} - \epsilon_t^{S,dw}\right) \quad (42)$$

$$\Delta_t^{vpp^{up}} \leq 2 \cdot \left(\overline{p}^w + \overline{p}^b\right) \cdot y_t^e \quad (43)$$

$$\Delta_t^{vpp^{dw}} \leq 2 \cdot \left(p^w + p^b\right) \cdot \left(1 - y_t^e\right) \quad (44)$$

$$soc_t^* = soc_{t-1}^* + \eta \cdot \left(e_t^{D,c*} + e_t^{S,dw*}\right) - \left(e_t^{D,d*} + e_t^{S,up*}\right) \quad (45)$$

$$\Delta_t^{soc} = soc_{t,1} - soc_t^* \quad (46)$$

The first term of (30) represents the revenue corresponding to the energy scheduled in the day-ahead electricity market by the GRM model, valued at the cleared market price $\pi_t^{D*}$. It is a constant term and could therefore be omitted. It is included in the objective function for the sake of clarity. The second and third terms represent the expected revenue and cost corresponding to the positive and negative energy imbalances $\Delta_t^{vpp^{up}}/\Delta_t^{vpp^{dw}}$. These energy imbalances account for the deviations with respect to both the energy scheduled by the GRM model in the day-ahead market and the secondary regulation energy required by the TSO in period $t$ and are valued at the upward/down imbalance prices according to the TSO rules. It is important to note that the variables $\Delta_t^{vpp^{up}}/\Delta_t^{vpp^{dw}}$ are not the same as the variables $\Delta_{t,s}^{up}/\Delta_{t,s}^{dw}$ used in the GRM model. The second line of (30) represents the economic result of the secondary regulation reserve market. The first two terms correspond to the secondary regulation reserve valued at the cleared market price whereas the third term represents the penalty in case of noncompliance of the secondary regulation reserve scheduled by the GRM model, according to [29]. The first two terms are constant and, therefore, could be omitted. They are included in the objective function for the sake of clarity. The third line of (30) represents the expected revenue corresponding to the secondary regulation energy upward and downward required by the TSO. Finally, the last line of (30) considers the battery degradation cost in the same way as the GRM model and, in addition, a term that monetises the battery state of charge at the end of the period $t$ with respect to the one scheduled by the GRM model, similarly to [30]. In the above description of (30) we use "expected revenue" to refer to those revenues which are calculated as a function of forecast prices (imbalance and secondary regulation energy prices). In the results presented in Section 4 these expected revenues are recalculated with the real prices in a post-process which uses the decisions provided by the EDM model.

The EDM model is subject to the deterministic version of the GRM model constraints (3)-(6), (17)-(18), (20)-(29). All variables in such constraints would be represented with the superscript $*$ and without considering the wind energy scenario $s$ in the formulation of the EDM model. These constraints are not included in the paper for the sake of conciseness.

In addition, the EDM model is subject to the constraints (31)-(46). The maximum power output of the wind farm is given by (31). Equations (32) and (33) ensure that the VPP's energy imbalance with respect to the energy scheduled in the day-ahead electricity market by the GRM model is positive or negative. The VPP's energy imbalance with respect to the energy scheduled in the day-ahead electricity market in hour $t \in T^1$ is given by (34).

The secondary regulation upward/downward reserve re-dispatched are calculated according to equations (35) and (36), respectively. Equations (37) and (38) are used to calculate the deviation with respect to the upward and downward reserve scheduled by the GRM model, respectively.

The secondary regulation upward/downward energy is calculated according to (39), where $\epsilon_t^{S,up}/\epsilon_t^{S,dw}$ represents the noncompliance of the upward/downward secondary regulation energy required by the TSO, respectively. The variable $\epsilon_t^{S,up}/\epsilon_t^{S,dw}$ is assigned a fixed null value when the right hand side of (39) is negative/positive. In addition, equations (40) and (41) include a binary variable $y^f$ that prevent the GRM model from providing upward secondary regulation energy in case the $\Delta_{D,t}^{vpp^{dw}}$ is positive. This operation might be interesting from a cross-arbitrage perspective but it is not possible since the existing measurement devices do not allow it.

The total VPP's energy deviation with respect to both the energy scheduled in the day-ahead market by the GRM model and the secondary regulation energy required by the TSO, is calculated according to (42). Equations (43) and (44) ensure that the total VPP's energy imbalance is positive or negative.

The battery state of charge $soc_t^*$ is calculated according to (45).

As mentioned above, the objective function (30) includes a term that economically values the deviation of the battery state of charge in period $t$ with respect to the one scheduled by the GRM model. This deviation is calculated according to (46). The methodology followed to monetise the deviation of the battery state of charge is similar to the one proposed in [30]: the state of charge deviation in period $t$, $\Delta_t^{soc}$, is multiplied by the parameter $soc^v$, which represents the VPP's storage opportunity cost given by GRM model. If the deviation is positive, there will be more energy stored in the battery than the scheduled one, and as a result, the value of the objective function will increase; if the deviation is negative, there will be less energy stored in the battery than the scheduled one, and consequently, the value of the objective function will decrease.

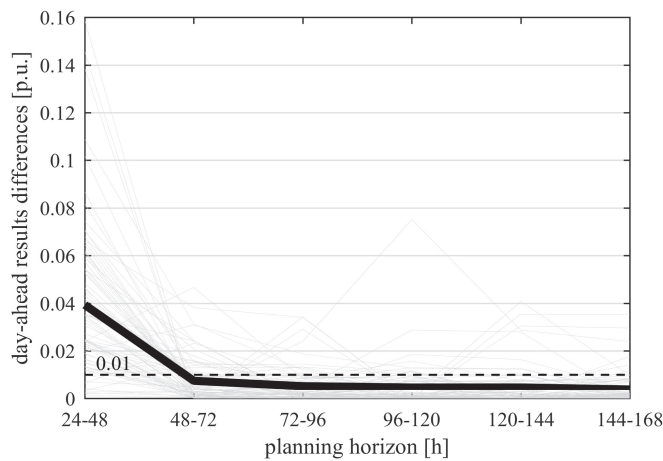

**Figure 4:** Absolute relative day-ahead VPP's result differences for different planning periods.

The battery degradation cost $c_t^{b*}$ is calculated in the same way as in the GRM model.

### 3.3. Parameter setting of the GRM model

In order to calculate the adequate planning period $T$ of the GRM model, a similar methodology to the one presented in [30] has been followed. A deterministic version of the GRM model has been used to analyse the sensitivity of its objective function to the length of the planning period (24-48-72-96-120-144-168 hours) over the set of 81 representative cases defined in Section 2.2. The absolute relative differences in the day-ahead VPP's result with different planning periods are depicted in Fig. 4 for the 81 cases. As can be seen in the figure, in most cases the results varies by less than 1% when using a planning period of 48 and 72 hours. Based on these results, a planning period of 72 hours is considered adequate.

Once the planning period $T$ of the GRM model has been determined, the appropriate number $m$ of wind energy scenarios $s$ to be considered in the GRM model must also be estimated. For this purpose, a methodology based on the silhouette criterion [31] has been applied to the available hourly wind generation dataset. For more details about this methodology, we refer the reader to [32]. From the application of this methodology, a value of $m = 3$ has shown to be enough to properly capture the wind energy variability in the next 72 hours. In addition to this methodology a scenario-based methodology [23] has been used for the same purpose. The result obtained with the silhouette criterion is consistent with the one obtained when using a scenario-based method.

## 4. Discussion of results

The GRM model and the EDM model have been applied in the above-mentioned set of 81 representative cases as depicted in Fig. 2. The execution of the models has been conducted on a computer with an Intel Xeon processor E5-2687@3.10 GHz and 64 GB RAM using the CPLEX commercial solver. A maximum computation time of 1,800 s with a target optimality gap of 1% have been set as stopping criteria in both models.

In order to highlight different aspects of the performance of the proposed models, different versions of the GRM model have been applied on the same set of cases as the GRM model, namely: a deterministic version of the GRM model which considers perfect knowledge of the input wind and market data (hereinafter referred to as DETPK), and a second deterministic version of the GRM model with imperfect knowledge (hereinafter referred to as DET), that considers the weighted average of the stochastic scenarios of the hourly wind energy generation considered in the GRM model.

Both the GRM model and the DET model have been run considering different forecast strategies of the percentage of the committed reserves (upward/downward) that are required in real-time, namely:

- C1: 100/0% of the committed reserve is required upward/downward in period $t$ according to [25],
- C2: 50/50% of the committed reserve is required upward/downward in period $t$,
- C3: the percentage of the committed reserve that is required upward/downward in real-time in period $t$ is equal to the average percentage of the committed reserve required in real-time in the last year in the same period.

The GRM model and the DET model with the above-mentioned forecast strategies will be hereinafter referred to as GRMC1, GRMC2, GRMC3 and DETC1, DETC2, DETC3 models, respectively. The results of the DETC* and GRMC* models are used as an input to the EDM model. The results of the DETPK model are not used as input to the EDM model since the DETPK model considers perfect knowledge of all wind and market data. Fig. 5 depicts the input data required when considering the different versions of the GRM model and the ones needed when using the EDM model.

The results presented in this section correspond to the ones given by the DETPK model, and the ones given by the EDM model using as inputs the results given by the GRMC* and DETC* models. The results provided by the EDM model are labelled in all figures and tables of this section using the name of the short-term scheduling model (DETC* or GRMC*) from which the input data have been taken.

Fig. 6 depicts the VPP's daily average income for each one of the above-mentioned models across the 81 cases. It is worth to mention that the revenue/cost due to both energy imbalances and the secondary regulation upward/downward energy given by EDM model has been considered with the cleared market price of those markets. To make a fair comparison, the daily income given by the EDM model with inputs from the DETC* and GRMC* models, have been corrected in a similar way to [23]: the differences observed in the battery's state of charge at hour $t=24$ are translated into monetary terms, using as conversion factor the VPP's storage opportunity cost $soc^v$ given by the DETPK model.

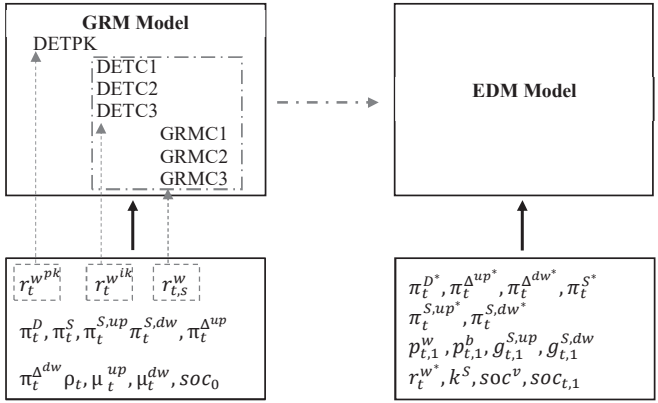

**Figure 5:** Data input of the optimisation models.

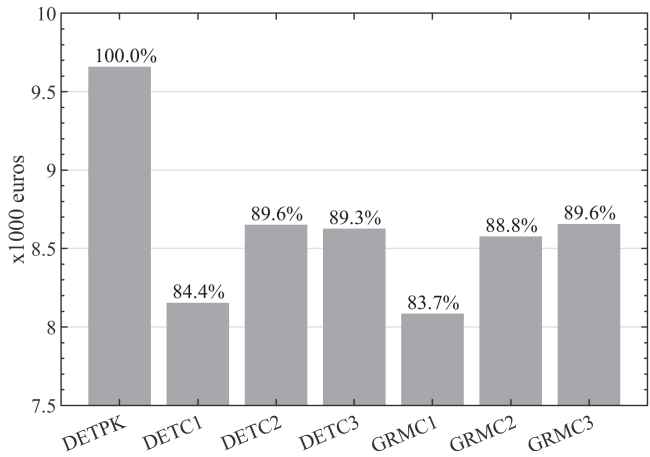

**Figure 6:** Daily average income of the VPP for each model considered.

As expected, the model that provides the best economic results is the one that assumes perfect knowledge, DETPK model, resulting a daily average income of 9,765 €. The models with imperfect knowledge that gives the highest daily income are the DETC2 and the GRMC3 model, i.e. the combination of the DET model and the forecast strategy C2 and of the GRM model and the forecast strategy C3. As can be seen in Fig. 6, the daily average income of the VPP when using these models, represents the 89.6% of the income obtained with perfect knowledge. As can be deduced from Fig. 6, forecast strategies C2 and C3 yield better results than forecast strategy C1. The difference between forecast strategies C2 and C3 is not significant.

Fig. 7 also depicts the VPP's daily average income for each one of the above-mentioned models across the 81 cases, highlighting in this case, the different sources of income or cost. The nomenclature used in the figure corresponds to the day-ahead electricity market (DEM), the energy imbalances (EI), the secondary regulation reserve (SRR), the non-compliance of the secondary regulation reserve (SRRD) and the secondary regulation energy (SRE). As expected, the results of the DETPK model account for no energy imbalance or non-compliance of the committed reserves, since it assumes perfect knowledge.

The income in the day-ahead electricity market obtained when using the DET* and GRMC* models is higher than the one obtained when considering perfect knowledge. However, this higher income comes at the expense of an also higher cost of energy imbalances. The DET* and GRMC* models do not only yield a high energy imbalance cost due to the noncompliance of the generation schedule of the day-ahead electricity market and of the secondary regulation energy required by the TSO (the cost of both imbalances are accounted for in EI), but also provide a moderate cost due to the noncompliance of the reserve schedule of the secondary regulation reserve market when using forecast strategies C2 and C3.

Another remarkable difference between the results provided by the DET* and GRMC* models can be found in Fig. 7. As can be seen in the figure the GRMC* models yield higher energy imbalance costs than the DET* models. This is mainly explained because, in general, the schedules given by the GRM* models have higher wind generation than the DET* ones. To illustrate this, Fig. 8 depicts the real wind generation together with the wind generation scheduled by both models for a single forecast strategy of the percentage of the committed reserves (C2) across the 81 cases. As can be seen in Fig. 8 the wind generation scheduled by the GRMC2 model is systematically higher than the real wind generation whereas the wind generation scheduled by the DETC2 model is more or less centered around the real wind generation.

No other remarkable differences between the models are found in Figs. 6 and 7.

The computational performance of the GRM model is summarised in Table 2. The parameters used to evaluate the the model's computational performance are the computational time and the optimality gap. As can be seen in Table 2, the performance of the model is sensitive to the considered forecast strategy of the percentage of the committed reserves (upward/downward) that are required in real-time, since the strategy has a strong impact on the energy management of the battery when solving the short-term scheduling problem. It is important to note that the times included in Table 2 only account for the computational time of the short-term scheduling. The computational time of each redispatch operation is less than 1 second in all cases.

As can be seen in Table 2 the models with imperfect knowledge which provide the highest average daily income,

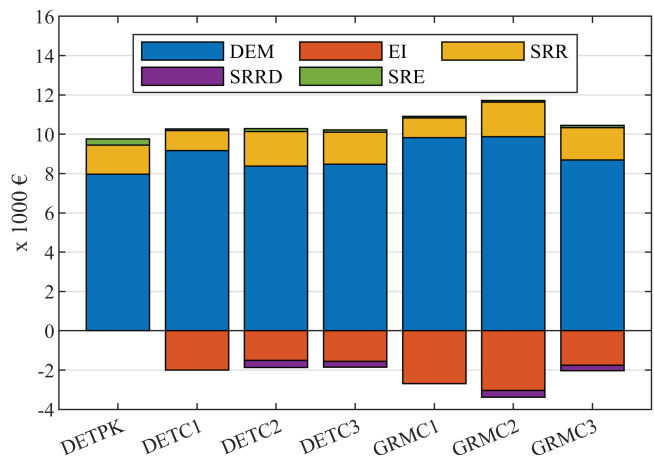

**Figure 7:** Daily average income or cost of the VPP for each market and model considered.

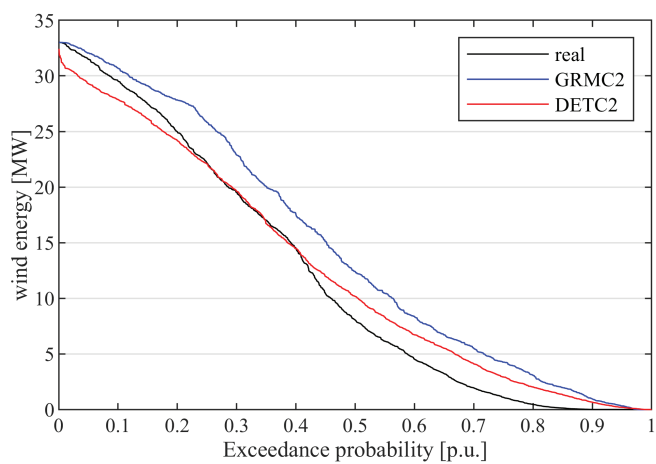

**Figure 8:** Wind generation scheduled by the GRMC2 and DETC2 models across the 81 cases.

**Table 2**
Computational performance results of DET/GRM models using different real-use second reserve criteria.

| Model | computational time [s] | | optimality gap [%] | |
|---|---|---|---|---|
| | mean | std | mean | std |
| DETC1 | 1717 | 369 | 2.18% | 0.91% |
| DETC2 | 1 | 1 | 0.15% | 0.10% |
| DETC3 | 14 | 20 | 0.93% | 0.12% |
| GRMC1 | 1759 | 263 | 2.70% | 1.04% |
| GRMC2 | 2 | 9 | 0.58% | 0.24% |
| GRMC3 | 452 | 640 | 1.01% | 0.11% |

DETC2 and GRMC3, have a good computational performance, as their mean optimality gap achieved is very low and their mean computational time is 1 and 452 seconds, respectively, making them fully practical for their daily use in the case study considered in this paper. Having in mind the standard deviation of the GRMC3 model's computational time, some cases might take between 15 and 30 minutes. This is still a reasonable time to schedule the generation and reserves of the considered VPP in the day-ahead electricity and secondary regulation reserve markets. However, the scalability of the GRMC3 model so that it can be used with larger VPPs comprising several renewable generation plants and energy storage units might be computationally hard. Even though the scalability of the models is out of the scope of the paper, we believe that the DETC2 model would be easily scalable to be used with larger VPPs.

Another version of the GRM and EDM models, but without considering the battery degradation cost, has been applied over the same set of cases analysed in this work. The results of these cases have been post-processed so as to compute the actual incurred battery degradation cost and to correct in a similar way to [23], the differences observed in the battery's state of charge at hour $t=24$. The incomes obtained when considering the battery degradation cost in the decision process are higher than those obtained when the cost is neglected by a factor ranging from 0.6% to 1.5%.

## 5. Conclusions

The results obtained in this work show that the optimisation models presented in the paper are effective tools for the short-term scheduling and redispatch of a virtual power plant (VPP) composed of a wind farm and a Li-ion battery, that participates in the day-ahead electricity and secondary regulation reserve markets of the Iberian electricity market.

The proposed optimisation models have been used in a realistic market setting to determine first the day-ahead generation and reserve schedule of a VPP and then the VPP's redispatch close to the time of energy delivery. The revenues obtained when using the proposed optimisation models are almost 90% of the ones obtained assuming perfect knowledge of all markets variables and of the available wind generation.

The results obtained in the paper show that the forecast strategy of the secondary regulation energy required in real-time by the TSO has a significant impact on the VPP's revenues and on the computational performance of the optimisation model used to determine the day-ahead VPP's generation and reserve schedule.

The computational performance of the optimisation model proposed to determine the day-ahead VPP's schedule is reasonable for its daily use. The application of the optimisation models proposed in this paper to other VPP configurations or market structures is straightforward. The deterministic version of the optimisation model proposed to determine the day-ahead VPP's schedule is easily scalable to be used with larger VPPs.

From the work carried out to develop the presented optimisation models and to obtain the results presented in this paper, we have identified three interesting research lines we are already working on. One is the use of a "dynamic" (i.e. updated hourly) storage opportunity cost in the VPP's redispatch, and others are assessing the economic viability of the battery's integration in an existing wind farm, assuming different power ratings and energy storage capacities, and upgrading the models to consider the intra-hourly variability of wind generation. Another future work will be the investigation of the optimal sizing of the battery for a VPP in the Iberian and other electricity markets.

## Acknowledgement

The work here presented has been funded by the Fundación Iberdrola under the project "Stochastic optimisation model for the short-term scheduling of generation and reserves of a Virtual Power Plant with wind generation and energy storage" awarded with a grant of the Call for Energy and Environment Research Grants 2019.